\documentclass[aps,amsmath,twocolumn,superscriptaddress,prl]{revtex4}
\usepackage{amsmath,amssymb,bm}
\usepackage[dvips]{graphicx,color}
\usepackage{braket}
\usepackage{ulem}
\usepackage{setspace}[10]
\usepackage{comment}
\usepackage[dvips]{color}
\usepackage[dvipsnames]{xcolor}
\usepackage{comment}
\usepackage{mathrsfs}

\newcommand{\ee}{\mathrm{e}}

\newcommand{\bfig}{\begin{figure}\begin{center}}
\newcommand{\efig}{\end{center}\end{figure}}

\begin{document}
\title{
Two-photon driven magnon-pair resonance as a signature of spin-nematic order}
\author{Masahiro Sato}
\email{masahiro.sato.phys@vc.ibaraki.ac.jp}
\affiliation{Department of Physics, Ibaraki University, Mito, 
Ibaraki 310-8512, Japan}
\author{Yoshitaka Morisaku}
\affiliation{Department of Physics, Ibaraki University, Mito, 
Ibaraki 310-8512, Japan}
\date{\today}
\begin{abstract}
We theoretically study the nonlinear magnetic resonance driven by intense laser or 
electromagnetic wave 
in a fully polarized frustrated magnet near a less-visible spin-nematic ordered phase. 
In general, both magnons and magnon pairs (two-magnon bound state) appear as the low-energy excitation 
in the saturated state of spin-nematic magnets. Their excitation energies are usually 
in the range between 10 gigahertz and 10 terahertz (THz). 
Magnon pairs with angular momentum 2$\hbar$ can be excited 
by the simultaneous absorption of two photons, and such multi-photon processes occur 
if the applied THz laser is strong enough. We compute laser-driven magnetic dynamics 
of a frustrated four-spin system with both magnon ($\hbar$) and magnon-pair (2$\hbar$) like excitations 
which is analogous to a macroscopic frustrated magnet with a spin nematic phase. 
We estimate the required strength of magnetic field of laser for the realization of two photon absorption, 
taking into account dissipation effects 
with the Lindblad equation. 
We show that intense THz laser with ac magnetic field of 0.1-1.0 Tesla is enough to observe magnon-pair resonance.  
\end{abstract}
\maketitle

{\it Introduction.}--
Laser science and technology have progressed in the last decades, and 
stimulated the study of condensed matter and nonequilibrium physics 
because the progress makes it possible to observe or create a variety of excitations in solids, liquids, and so on. 
In recent years, 
the laser science in the regime of 0.1-10 terahertz (THz)~\cite{Hirori11,Sato13,Dhillon17,Liu17} 
has strikingly developed and we can use THz laser pulses with intensity of 1 MV/cm ($\sim$0.3 Tesla). 
As a result, it is becoming possible to control magnetic excitations or textures with the laser 
because the photon energy in the THz range is comparable to that of magnetic excitations, 
especially, those of antiferromagnets~\cite{Kimel18}. Photo-induced magnetic phenomena have also been actively 
explored as the issues of magneto-optics~\cite{Kirilyuk10} and spintronics~\cite{Maekawa12}. 
Several groups have observed linear and nonlinear magnetic responses for THz laser or waves: 
For instance, large magnetic resonances driven by THz laser or wave~\cite{Mukai16,Baierl16}, 
high harmonic generation (HHG) induced by a THz laser pulse in an antiferromagnetic insulator~\cite{Lu17}, 
electro-magnon resonance driven by an electric-field of THz wave~\cite{Pimenov06,Takahashi11,Staub14}, 
dichroisms in a ferrimagnet driven by THz vortex beam~\cite{Sirenko19}, etc. 
In addition, the ESR driven by electromagnetic waves in the range between 10 gigahertz (GHz) 
and 1.0 THz has been long 
studied~\cite{Slichter,Kubo54,Mori62,Richards74,Oshikawa99,Oshikawa02,Furuya15,Nojiri03,Tanaka03,Smirnov12,Zvyagin15}. 
Microscopic or quantum theories for magnetic dynamics driven by intense electromagnetic waves 
have also begun to develop: Floquet engineering in magnetic systems~\cite{Takayoshi14-1,Takayoshi14-2,Sato14,Sato16,Higashikawa18}, 
control of exchange couplings in Mott insulators~\cite{Mentink15,Takasan19}, 
ultrafast creation or control of magnetic defects in chiral magnets~\cite{Mochizuki10,Koshibae14,Fujita17-1,Fujita17-2}, 
applications of topological light waves~\cite{Fujita18,Fujita19}, laser-driven spin current in magnetic insulators~\cite{Ishizuka19-1,Ishizuka19-2}, 
HHG in quantum spin systems~\cite{Ikeda19}, etc.

\begin{figure}[t]
\begin{center}
\includegraphics[width=8cm]{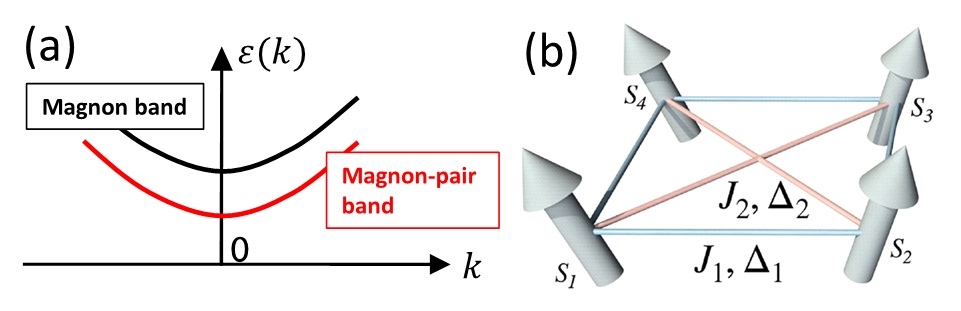}
\caption{(Color online) (a) Generic band structures of magnon and magnon pair 
in the field-driven fully polarized state of a frustrated magnet near a spin nematic phase~\cite{Supple}. 
(b) Frustrated four-spin model [see Eq.~(\ref{eq:Model})].}
\label{fig:Band_Model}
\end{center}
\end{figure}
Motivated by these activities, 
in this paper, we theoretically consider how to detect a signature of less-visible spin nematic (quadrupolar) order 
in magnetic insulators with laser or electromagnetic wave. 
Spin nematic ordered phase~\cite{Chubukov91,Penc06,Shannon06} is the physical state with a spin quadrupolar order and 
without any spin dipolar (magnetic) order. Its order parameter is defined as an expectation value of tensor product of two spins. 
In the present work, we focus on the spin nematic order on $S^x$-$S^y$ plane defined by 
$\langle S^+_{\bm r}S^+_{\bm r'}+ S^-_{\bm r}S^-_{\bm r'}\rangle$, which accompanies 
the breaking of U(1) spin rotation symmetry around the $S^z$ axis. 
It is important to inhibit usual spin order for the emergence of such spin nematic states and 
thereby frustrated magnets often become its nice candidate. 
Another point is that not only standard magnons but also magnon pairs 
(molecules of two magnons)~\cite{Kecke07,Ueda09,Zhitomirsky10} 
usually appear in field-induced fully polarized (i.e., ferromagnetic) states of spin nematic magnets 
[See Fig.~\ref{fig:Band_Model}(a) and Supplement Material~\cite{Supple}]. 
If the applied magnetic field is decreased and the magnon-pair band becomes lower than 
the energy of the saturated state, the Bose Einstein condensation (BEC) of magnon pairs occurs. 
The product of neighboring spins $S^-_{\bm r}S^-_{\bm r'}$ ($S^+_{\bm r}S^+_{\bm r'}$) can be viewed as 
the creation (annihilation) operator of a magnon pair. In their BEC state, 
these operators have finite expectation values $\langle S^\pm_{\bm r}S^\pm_{\bm r'}\rangle\neq 0$ 
and therefore it means the emergence of a spin nematic order.       
This is a typical scenario of generating a spin nematic order.

Generally, it is quite difficult to detect a clear evidence for the spin nematic order compared to usual magnetic orders 
because its detection requires a direct observation of a tensor product of two spins 
$\langle S^+_{\bm r}S^+_{\bm r'}+ S^-_{\bm r}S^-_{\bm r'}\rangle$
or a four-point spin-nematic correlation function such as 
$\langle S^+_{\bm r}S^+_{\bm r+\bm \delta} S^-_{\bm 0}S^-_{\bm \delta}\rangle$. 
For the spin-nematic quasi long-range ordered phase in one-dimensional magnets, 
it has been shown~\cite{Sato09,Sato11} that NMR~\cite{Nawa13,Buttgen14,Matsui17,Nawa17,Grafe17}, 
neutron scattering spectra~\cite{Masuda11,Mourigal12}, and spin Seebeck effect~\cite{Hirobe19} 
are very useful to detect its signature, 
while clear experimental ways of detecting spin-nematic long-range orders 
have not been well established~\cite{Shindou13,Smerald13,Furuya18}.    
On the other hand, as we mentioned above, magnon-pair excitations almost always appear in the saturated state of 
spin-nematic magnets including both spin-neamtic long-range and quasi long-range ordered phases. 
We here discuss a method of observing magnon pairs with intense laser or electromagnetic waves 
as a way of obtaining an indirect but strong evidence for spin nematic orders. 
Magnons and photons carry angular momentum $\hbar$, while magnon pairs have angular momentum $2\hbar$. 
Therefore, magnon pairs can be excited through two photon absorption and 
such multi-photon processes can be realized with sufficiently strong laser.  
We compute the time evolution of laser-driven spin dynamics in a frustrated nano spin model 
that is a mimicry of spin nematic magnets. 
We take into account the dissipation effect, 
which is quite important to estimate realistic spectra of magnon-pair resonance, 
by applying quantum master equation with Lindblad approximation.   
We show that magnon-pair resonance spectra can be detected 
with currently available THz laser or gigahertz (GHz) wave.

{\it Model and method.}--
Here we define our model for studying the laser-driven spin dynamics. 
We focus on the frustrated four-spin model described by Fig.~\ref{fig:Band_Model} (b). 
The Hamiltonian is given by 
\begin{align}
{\cal H}_0=&\sum_{j=1-4}\left(J_1{\bm S}_j\cdot{\bm S}_{j+1} + \varDelta_1{S}_j^z{S}_{j+1}^z\right) 
-H S_{\rm tot}^z\notag\\
  &+\sum_{j=1,2}\left(J_2{\bm S}_j\cdot{\bm S}_{j+2} + \varDelta_2{S}_j^z{S}_{j+2}^z	\right),
\label{eq:Model}
\end{align}
where $\bm S_j$ is the electron spin-$\frac{1}2$ operator on $j$-th site ($j$: mod 4), and 
$S_{\rm tot}^\alpha=\sum_{j=1}^4 S_j^\alpha$ is the sum of four spins.
Here, $J_{1,2}$ are the competing exchange interactions, $\Delta_{1,2}$ are the Ising anisotropy constants, and 
$H=g\mu_B h_0$ is the strength of Zeeman coupling for an applied static magnetic field $h_0$ 
($g$ is g factor and $\mu_B$ is Bohr magneton). This paper uses the unit of $\hbar=1$. 
The eigen energies and normalized eigenstates for ${\cal H}_0$
are respectively described as $\{E_n\}$ and $\{|\psi_n\rangle\}$ with $E_1\leq E_2\leq\cdots\leq E_{16}$. 
It is shown that two-dimensional system consisting of weakly coupled four-spin models~(\ref{eq:Model}) 
exhibits a spin nematic order at $J_1/J_2\sim -2$~\cite{Ueda07}. The model~(\ref{eq:Model}) thereby 
may be regarded as a simple mimicry of a bulk spin-nematic magnet. 
Hereafter, we adopt $J_1=-2.2$, $J_2=1$ and $\Delta_1=\Delta_2=0.24$, in which 
the energy eigenstates are classified as a spin quintet $|S_{\rm tot},S_{\rm tot}^z\rangle=|2,M\rangle$, 
three spin triplets $|1,M\rangle_p$, and two spin singlets $|0,0\rangle_q$ ($p=1,2,3$ and $q=1,2$).  
As we show later, a finite anisotropy $\Delta_{1,2}$ generates a difference between resonant frequencies 
of magnon and magnon pair. 
The details of $\{|\psi_n\rangle\}$ and 
$\{E_n\}$ are explained in Supplementary Material~\cite{Supple}. 

Figure~\ref{fig:Level} (a) shows the field dependence of energy levels in the low-energy range. 
Since we consider the fully polarized state and magnon-pair like excitations in the model~(\ref{eq:Model}), 
we set $H=0.56$, in which the ground state is fully polarized ($S^z_{\rm tot}=2$) as shown in Fig.~\ref{fig:Level} (a). 
Then we calculate the spin dynamics under the application of intense THz laser or electromagnetic wave.  
The ac Zeeman coupling is given by 
${\cal H}_{\rm cp}(t)=\frac{A}{2}(e^{-i\omega t}S^+_{\rm tot}+e^{+i\omega t}S^-_{\rm tot})$ 
for a circularly polarized laser and 
${\cal H}_{\rm lp}(t)=A\cos(\omega t)S_{\rm tot}^x=\frac{A}{2}\cos(\omega t)(S^+_{\rm tot}+S^-_{\rm tot})$ 
for a linearly polarized one. $\omega$ is the angular frequency of laser, 
$A=g\mu_B h_{\rm ac}$ denotes the amplitude of magnetic field of laser and 
$S^{\pm}_{\rm tot}=S_{\rm tot}^x\pm iS_{\rm tot}^y$. 
The ac electric field is negligible 
since the THz photon energy is usually much smaller than the charge gap of magnets.

\begin{figure}[t]
\begin{center}
\includegraphics[width=8cm]{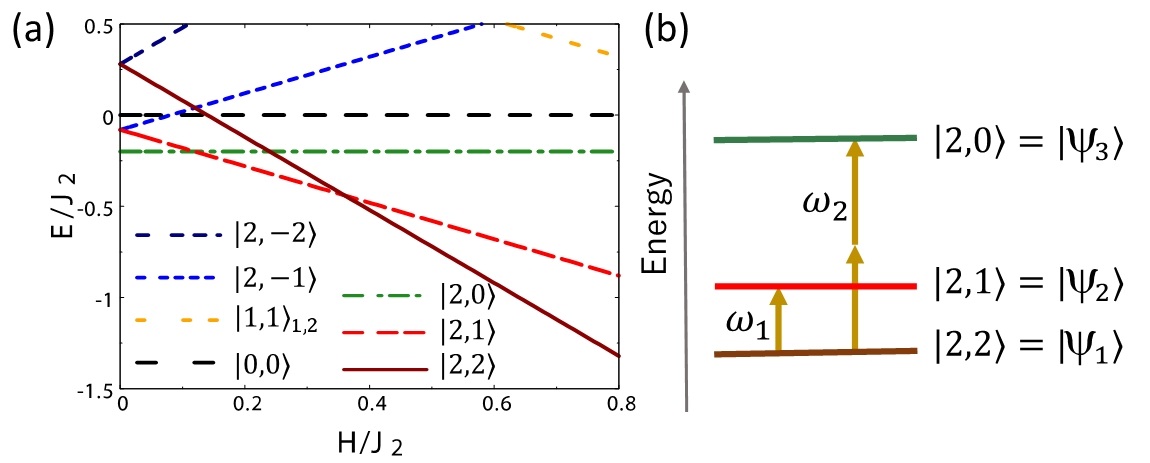}
\caption{(Color online) (a) Field ($H$) dependence of energy levels $|S_{\rm tot}, M\rangle$ 
of the nano spin model~(\ref{eq:Model}) with $J_1=-2.2$, $J_2=1$, and $\Delta_1=\Delta_2=0.24$. 
(b) Low-energy level structure at a high field $H=0.56$. 
$\omega_{1(2)}$ is the magnon (magnon pair) resonance frequency. 
The unit of $\hbar=1$ is used.}
\label{fig:Level}
\end{center}
\end{figure}

To see the time evolution of the nano magnet with ac Zeeman coupling, we numerically solve 
the quantum master equation for density matrix $\rho(t)$~\cite{Breuer,Lindblad,GKS,Ikeda20},
\begin{align}\label{eq:master}
	\dot\rho(t)=&-i[{\cal H}(t),\rho(t)]\notag\\
&+\sum_{j=2}^{16}[N(\omega_j)+1]\left(L_j\rho(t)L_j^\dagger-\frac{1}{2}\left\{L^\dagger_j L_j,\rho(t)\right\}\right)\notag\\
		&+\sum_{j=2}^{16}N(\omega_j)\left(L_j^\dagger\rho(t)L_j-\frac{1}{2}\left\{L_j L_j^\dagger,\rho(t)\right\}\right),
\end{align}
with the fourth-order Runge-Kutta method. 
The first line of the r.h.s. represents the dynamics driven by the Hamiltonian ${\cal H}(t)={\cal H}_0+{\cal H}_{\rm cp}(t)$ 
(or ${\cal H}_0+{\cal H}_{\rm lp}(t)$), while the second and third lines give a so-called Lindblad type dissipation.   
The Lindblad (jump) operator $L_j=\sqrt{\gamma}|\psi_{j-1}\rangle\langle \psi_j|$ and 
$\gamma$ is the coupling constant between the system and environment. 
The value of $\hbar/\gamma$ is the typical time of relaxation. 
If there is a degeneracy $E_{j-1}=E_j$, we modify the jump operators as 
$L_{j-1}=\sqrt{\gamma/2}|\psi_{j-2}\rangle(\langle\psi_{j-1}|+\langle \psi_{j}|)$, 
$L_j=0$ and $L_{j+1}=\sqrt{\gamma/2}(|\psi_{j-1}\rangle+|\psi_{j}\rangle)\langle \psi_{j+1}|$. 
We set $N(\omega_j)=1/(e^{\omega_j/(k_BT)}-1)$ with $\omega_j=E_j-E_{j-1}$ so that the system relaxes to 
the equilibrium state of ${\cal H}_0$ at temperature $T$.

A many-spin model with a spin nematic phase is surely superior to the nano spin model~(\ref{eq:Model}) 
for the purpose of studying magnon-pair resonance. 
However, there are at least three reasons why the model~(\ref{eq:Model}) is expected to capture 
the essential aspect of a magnon-pair resonance in bulk systems. 
Firstly, the diffraction limit ($\sim$ wave length) of THz laser is much larger than the lattice space of magnets 
and therefore only magnetic excitations around wave number $\bm k=\bm 0$ are relevant for laser application. 
Even bulk magnets have only a few discrete modes around $\bm k=\bm 0$~\cite{Nojiri03,Tanaka03,Smirnov12,Zvyagin15} and 
excited states in the model~(\ref{eq:Model}) may be viewed as analogs of these $\bm k=\bm 0$ modes. 
Secondly, positions of two magnons in a single magnon pair are quite close to each other~\cite{Shannon06,Kecke07} 
because the attractive force between two magnons stems from short-range exchanges. 
Therefore, excited states with two (one) down spins in the model~(\ref{eq:Model}) are analogous 
to those with a magnon-pair (magnon) in a bulk magnet. 
The third point, which is most important, is that one can practically take the dissipation effect into account 
within the Lindblad approximation if the spin system is small enough. 
For the analysis of realistic magnetic-resonance spectra, 
small interactions breaking spin conservation and the spin-bath coupling 
(e.g., magnetic anisotropies, dipole interaction, spin-phonon coupling, etc.) are important rather than many-body effects. 
In fact, observed ESR spectrum shapes of nano magnets~\cite{Barra97,Nojiri06,Nojiri12} 
are often similar to those of bulk magnets~\cite{Nojiri03,Tanaka03,Smirnov12,Zvyagin15}. 
The Lindblad term phenomenologically describes the effect of such small interactions 
and makes the system relax to the equilibrium state. 
For correlated many-spin systems, even finding energy eigenstates is difficult and 
treating the dissipation effect in such bulk systems is a massively hard task. Moreover, if we continuously apply laser to 
an isolated many-spin system decoupled to environment, the system is generally heated up.   
From these arguments, we discuss a magnon-pair resonance by using the model~(\ref{eq:Model}) 
with the quantum master equation~(\ref{eq:master}). 
The analysis of spin dynamics in dissipative {\it many-spin} systems is left to a feature issue.

We note that the magnon-pair band is sometimes located around wave number $k^\alpha=\pi$ ($\alpha=x$, $y$, or $z$) 
in {\it antiferromagnetic} spin-nematic magnets~\cite{Shannon06,Kecke07,Ueda09,Zhitomirsky10,Sato09,Sato11}. 
Magnon pairs on such a band seem not to be coupled to 
applied THz or GHz electromagnetic waves. However, even in those cases, if the crystal symmetry is low enough and 
the unit cell includes multiple magnetic ions (e.g., due to dimerization), 
the usual ac Zeeman coupling of THz laser can excite magnon pairs. 
In addition, if a magnetoelectric coupling~\cite{Tokura14} exists in the spin-nematic magnets, 
excitations around $k^\alpha=\pi$ can be often created with laser~\cite{Pimenov06,Takahashi11,Staub14,Sato14,Sato16}.

{\it Analysis and results.}--
Based on the master equation~(\ref{eq:master}), 
we study laser-driven magnetic resonance in the model~(\ref{eq:Model}) with 
$J_1=-2.2$, $J_2=1$, $\Delta_1=\Delta_2=0.24$, and $H=0.56$. 
We consider the low temperature range of $k_BT\alt 0.1$. 
The initial state at $t=0$ is set to be a polarized equilibrium state with a temperature $T$ 
and then we add the ac Zeeman coupling ${\cal H}_{\rm cp}$ or ${\cal H}_{\rm lp}$. 
The Lindblad term helps the laser-driven system to return to the equilibrium state at $T$. 
From the low-energy levels of Eq.~(\ref{eq:Model}) shown in Fig.~\ref{fig:Level}, 
one sees that $|\psi_1\rangle$, $|\psi_2\rangle$, and $|\psi_3\rangle$
have $S^z_{\rm tot}=2$, 1, and 0, respectively. Therefore, we may view 
$|\psi_2\rangle$ and $|\psi_3\rangle$ as 
magnon and magnon-pair states, respectively.  
As we mentioned, magnons (magnon-pairs) can be resonantly excited by single photon (two photons).  
Thus the frequency of the magnon resonance is given by $\omega_1=E_2-E_1=H-1.5\Delta_1=0.2$, while 
that of magnon-pair resonance is $\omega_2=(E_3-E_1)/2=H-\Delta_1=0.32$, as shown in Fig.~\ref{fig:Level}(b). 
We will consider the range of laser frequency $\omega$ including 
these resonant values $\omega_1$ and $\omega_2$.  
We note that $\omega_2-\omega_1=0.12$ is much smaller than $J_2=1$ and 
it means that our set up imposes a tough condition to distinguish two resonance peaks. 

\begin{figure}[t]
\begin{center}
\includegraphics[width=8cm]{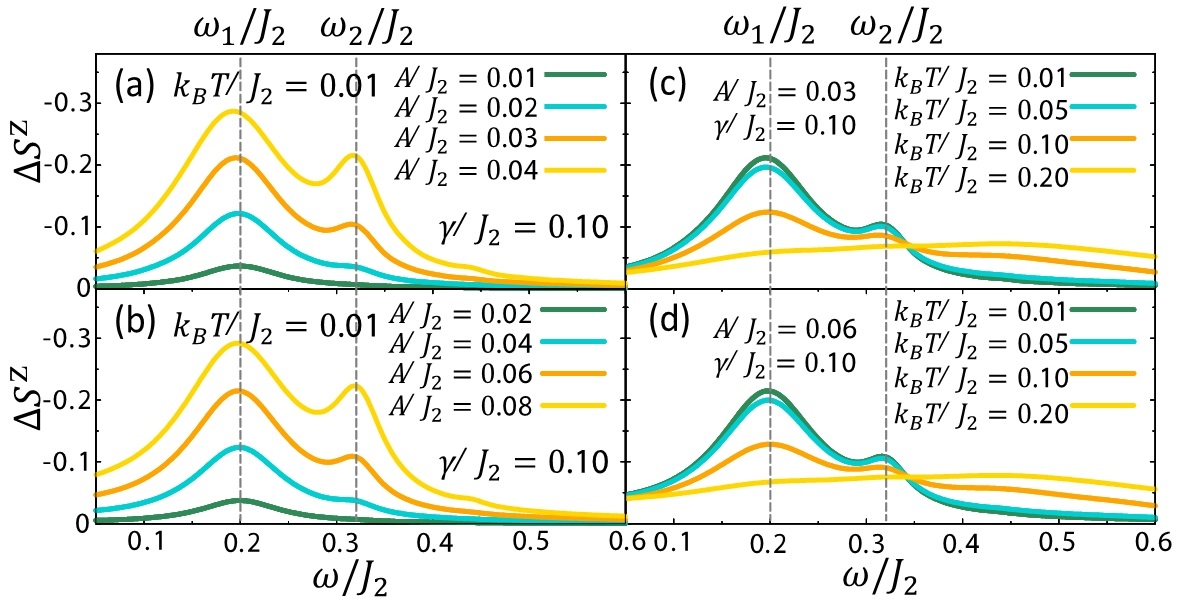}
\caption{(Color online) Spectra $\Delta S^z(\omega)$ for different values of laser strength or temperature 
in the case of circularly polarized laser (a)(c) and linearly polarized laser (b)(d). 
Parameters are set to be $J_1=-2.2$, $J_2=1$, $\Delta_1=\Delta_2=0.24$, and $H=0.56$. 
Dotted lines denote the resonant positions at $\omega=\omega_1$ and $\omega_2$.}
\label{fig:Spectra}
\end{center}
\begin{center}
\includegraphics[width=8cm]{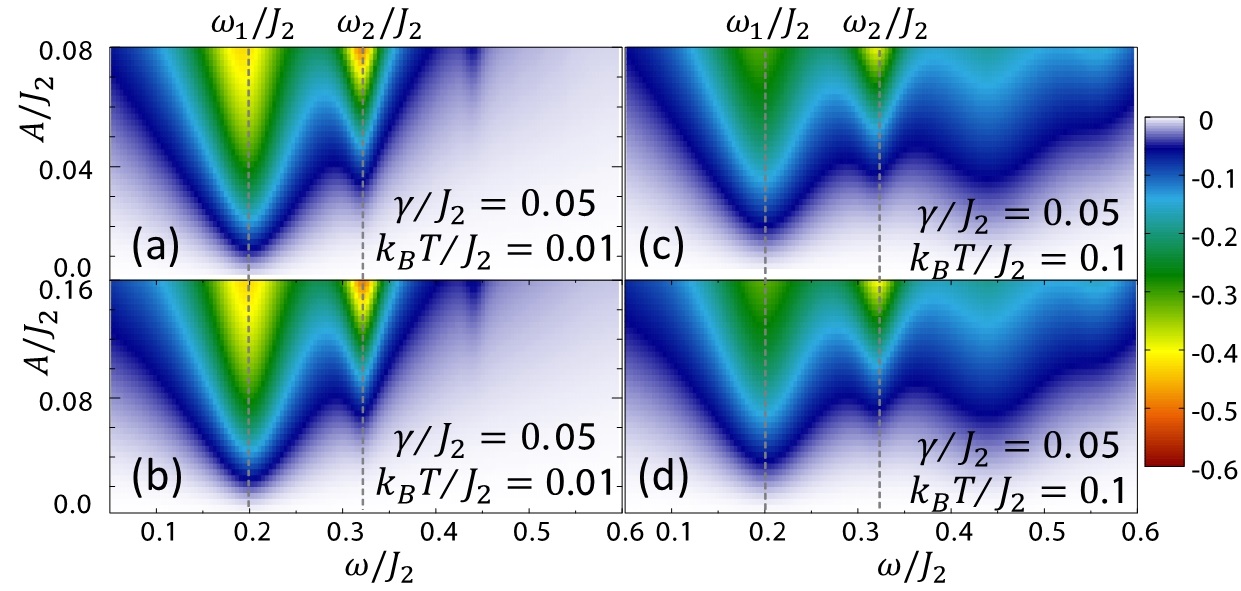}
\caption{(Color online) Laser-strength $A$ and frequency $\omega$ dependences 
of the spectra $\Delta S^z(\omega)$ in the case of circularly polarized laser (a)(c) 
and linearly polarized one (b)(d). 
We set $J_1=-2.2$, $J_2=1$, $\Delta_1=\Delta_2=0.24$, and $H=0.56$. 
Dotted lines denote the resonant positions at $\omega=\omega_1$, and $\omega_2$. 
}
\label{fig:Color}
\end{center}
\end{figure}

One can numerically calculate the expectation value of any operator ${\cal O}$ at arbitrary time $t$ 
from the density matrix $\rho(t)$: $\langle {\cal O}(t)\rangle={\rm Tr}[{\cal O}\rho(t)]$. 
Here, we concentrate on the magnetization change between the initial state and 
the nonequilibrium steady one~\cite{Ikeda20} 
which is realized by waiting for a long time from the beginning of laser application.   
Namely, we compute $\Delta S^z=\langle S^z_{\rm tot}\rangle_{\rm neq}-\langle S^z_{\rm tot}\rangle_{\rm eq}$, 
where 
\begin{align}\label{eq:Change}
	\langle S^z_{\rm tot}\rangle_{\rm neq}&= \frac{1}{\tau_0}\int_{\tau}^{\tau+\tau_0} dt \langle S^z_{\rm tot}(t)\rangle
\end{align}
and $\langle S^z_{\rm tot}\rangle_{\rm eq}$ is the initial value of $\langle S^z_{\rm tot}\rangle$ at a fixed $k_BT$. 
Here, $\tau$ and $\tau_0$ are set to be sufficiently larger than the relaxation time $\hbar/\gamma$ and 
the period $2\pi/\omega$ of laser, respectively. The integration of Eq.~(\ref{eq:Change}) is necessary 
to eliminate the small fluctuation of $\langle S^z_{\rm tot}(t)\rangle$, 
especially, in the case of linearly polarized laser. 
A large $|\Delta S^z|$ indicates a large precession motion (i.e., a large oscillation of transverse magnetization) 
driven by laser~\cite{Slichter} and it means that the system efficiently absorbs photons. 
Therefore, we can use $|\Delta S^z|$ as an index of the observability of magnon-pair resonances.

The relaxation time of electron spins in solids is usually 
from pico to nano seconds~\cite{Oshikawa99,Oshikawa02,Furuya15,Kirilyuk10,Beaurepaire96,Koopmans00,Mashkovich19,Tzschaschel19,Lenz06,Vittoria10}.
The current THz-laser technique enables us to utilize intense THz laser pulses 
with a few Tesla, which corresponds to a few MV/cm~\cite{Hirori11,Sato13,Dhillon17,Liu17}.  
For the reality, 
we hence consider the range of $\gamma\sim 0.01J_2-0.1 J_2$ and $A/J_2\alt 0.1$ 
in the numerical calculation of $\Delta S^z$: 
For instance, for $J_2/k_B=10$ [K] (50 [K]), $\gamma=0.05J_2$ and $A=0.05J_2$ respectively 
correspond to the relaxation time $\hbar/\gamma \simeq 15.2$ [ps] (3.0 [ps]) and 
the ac magnetic field $h_{\rm ac}\simeq 0.37$ (1.86) Tesla.

Figure~\ref{fig:Spectra} depicts computed $\Delta S^z(\omega)$ as a function of $\omega$, 
changing the laser strength $A$ or temperature $T$. We find that when the laser is weak in low $T$, 
only the magnon resonance at $\omega=\omega_1$ is clearly observed 
(this corresponds to the standard magnetic resonance), 
while magnon-pair peaks gradually grow up with increase of $A$. 
One sees that magnon-pair peaks become visible for $A\agt 0.03J_2$ ($A\agt 0.06J_2$) 
in the case of circularly (linearly) polarized laser. 
Therefore the result of Fig.~\ref{fig:Spectra} indicates that the magnon-pair resonance can be observed 
with an available strong laser or electromagnetic wave. 
Figure~\ref{fig:Color} shows the laser-induced magnetization $\Delta S^z$ 
for both circularly- and linearly-polarized lasers in a large range of $(A,\omega)$. 
This figure also tells us that magnon-pair peaks at $\omega=\omega_2$ become visible 
if the applied wave is strong enough.

\begin{figure}[t]
\begin{center}
\includegraphics[width=7cm]{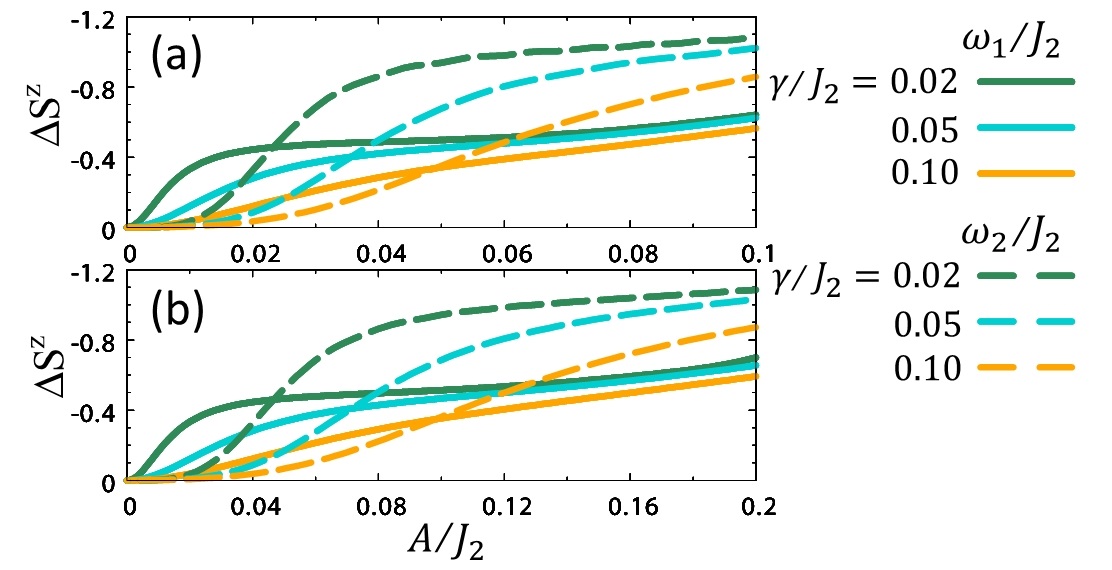}
\caption{(Color online) Laser strength dependence of the difference $\Delta S^z(\omega)$ 
at the resonant points $\omega=\omega_1$ and $\omega_2$ 
for (a) circularly and (b) linearly polarized laser. We use the parameters 
$J_1=-2.2$, $J_2=1$, $\Delta_1=\Delta_2=0.24$, $H=0.56$, and $k_BT=0.01$. }
\label{fig:Nonlinear}
\end{center}
\end{figure}

To more quantitatively see the required intensity of the laser, 
we show the laser-strength dependence of $\Delta S^z$ at the resonant points 
$\omega=\omega_1$ and $\omega_2$ in Fig.~\ref{fig:Nonlinear}. 
We find that the magnon-resonance peak almost linearly increases with $A$, 
especially, in a weak dissipation regime, 
whereas the magnon-pair peak exhibits a nonlinear increase in terms of $A$. 
Moreover, the magnon-pair peak becomes comparable to the magnon one if $A$ is sufficiently strong 
($A\agt 0.01J_2-0.05 J_2$).

\begin{figure}[t]
\begin{center}
\includegraphics[width=8cm]{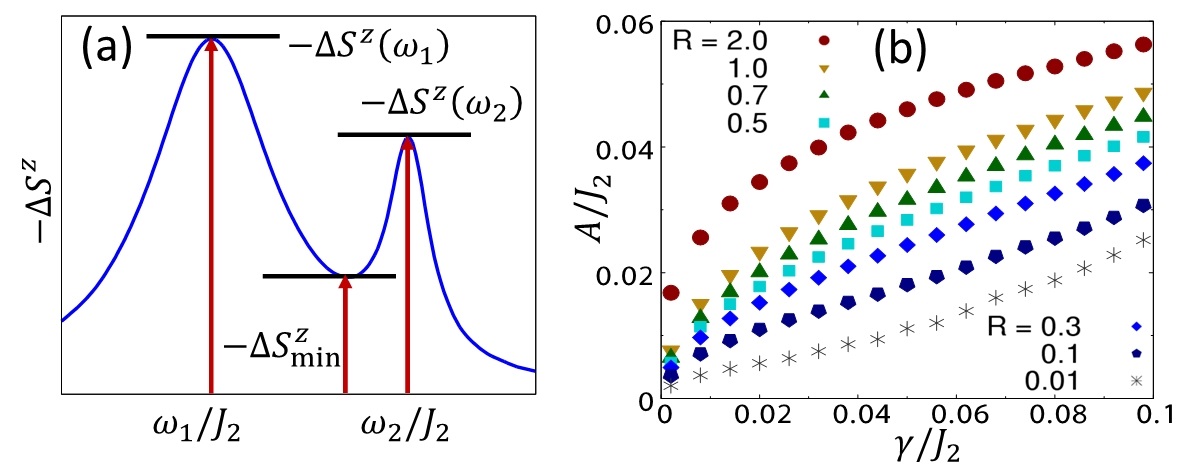}
\caption{(Color online) (a) Definition of $\Delta S^z_{\rm min}$. 
(b) Contour line of the ratio $R$ of Eq.~(\ref{eq:Ratio}) in $(\gamma/J_2,A/J_2)$ space 
in the case of circularly polarized laser. 
We again set $J_1=-2.2$, $J_2=1$, $\Delta_1=\Delta_2=0.24$, $H=0.56$, and $k_BT=0.01$. 
We also obtain a similar contour line in the case of linearly polarized laser.}
\label{fig:Ratio}
\end{center}
\end{figure}
Finally, we introduce another index for the visibility of magnon-pair resonances. 
As shown in Fig.~\ref{fig:Ratio}(a), we first define $-\Delta S_{\rm min}^z$ 
as the minimum value of $|\Delta S^z(\omega)|$ in the range between $\omega=\omega_1$ and $\omega_2$. 
Using it, let us consider the following quantity
\begin{align}\label{eq:Ratio}
	R(\omega_1,\omega_2)= 
\frac{\Delta S^z(\omega_2)-\Delta S_{\rm min}^z}{\Delta S^z(\omega_1)-\Delta S_{\rm min}^z}.
\end{align}
A sufficient large $R$ means a high possibility of detecting the magnon-pair peak in real experiment. 
Figure~\ref{fig:Ratio}(b) draws the contour curve of $R$ in $(\gamma, A)$ space, 
and it clearly indicates that a large laser strength $A$ and a small dissipation constant $\gamma$ are 
better for the observation of magnon-pair resonance.  
If we assume that $R>0.1$ is the necessary condition about the observation of the magnon-pair peak, 
Fig.~\ref{fig:Ratio}(b) implies that the peak can be observed in the region of $A\agt 0.3\gamma$.  

From Figs.~\ref{fig:Spectra}-\ref{fig:Ratio}, we conclude that 
one can observe not only magnon but also magnon-pair resonances in fully-polarized states of 
spin-nematic magnets 
if the laser intensity reaches $h_{\rm ac}\sim 0.1-1.0$ Tesla and 
the resonant point $\omega_{1,2}$ are sufficiently separated.

{\it Conclusions.}--
In summary, we theoretically discussed the observability of magnon-pair resonance 
in fully-polarized states of spin-nematic magnets. 
We compute the laser-driven spin dynamics in the frustrated nano spin model of Eq.~(\ref{eq:Model}), 
which is analogous to a bulk spin-nematic magnet, by applying the Lindblad equation. 
Our calculation strongly indicates that a currently available intense laser with $h_{\rm ac}\alt 1$ Tesla 
is enough to observe magnon-pair resonances. 
Besides spin-nematic or nano magnets, bound states of magnons also emerge in a class of 
frustrated or low-dimensional quantum magnets~\cite{Frustration11,Gogolin08}. 
Our estimation of the required laser strength for magnon-pair resonances 
would be applicable to such magnets.

\begin{acknowledgments}
We thank Satoshi Aizawa, Nobuo Furukawa, and Yusuke Yoshimoto for the discussion 
at the early stage of the present study. We also thank Y. Yoshimoto 
for drawing Fig.~\ref{fig:Band_Model}(b) and Shunsuke C. Furuya for the discussion about ESR. 
M. S. is supported by 
JSPS KAKENHI (Grant No. 17K05513 and No. 20H01830) and 
a Grant-in-Aid for Scientific Research on Innovative Areas
``Quantum Liquid Crystals'' (Grant No. JP19H05825).  
\end{acknowledgments}


\clearpage
\setcounter{figure}{0}
\setcounter{equation}{0}
\setcounter{section}{0}

\onecolumngrid

\begin{center}

\vspace{1.5cm}

{\large \bf Supplemental Material:\\
Two-photon driven magnon-pair resonance as a signature of spin-nematic order}

\vspace{0.3cm}

{\large Masahiro Sato$^{1}$ and Yoshitaka Morisaku$^{2}$} \\[2mm]
$^{1,2}$\textit{Department of Physics, Ibaraki University, Mito, Ibaraki 310-8512, Japan}

\end{center}

\renewcommand{\theequation}{S\arabic{equation}}
\renewcommand{\thesection}{S\arabic{section}}
\renewcommand{\thefigure}{S\arabic{figure}}
\renewcommand{\thetable}{S\arabic{table}}

\makeatletter
 \renewcommand{\@cite}[2]{[S#1]}
 \renewcommand{\@biblabel}[1]{[S#1]}

\section{Fully Polarized States of Spin Nematic Magnets}
Here, we shortly review a few universal features of field-driven fully polarized states in spin-nematic magnets. 
As we mentioned in the main text, in addition to usual magnons, 
magnon-pair excitations (two-magnon bound states) generally appear 
in the fully polarized states of a broad class of spin-nematic magnets. 
For simplicity, we assume that both magnon and magnon-pair bands are located around 
the wave-number vector $\bm k=\bm 0$, as shown in Fig. 1(a) of the main text. 
In the vicinity of the band bottoms, magnon and magnon-pair bands, $\epsilon_1(\bm k)$ and $\epsilon_2(\bm k)$, 
can be approximated by  
\begin{align}
\label{eq:bands}
\epsilon_1(\bm k) &= A_1 \bm k^2+g\mu_{\rm B} \hbar (h-h_{c1}),\\
\epsilon_2(\bm k) &= A_2 \bm k^2+2g\mu_{\rm B} \hbar (h-h_{c2}),
\end{align}
where $g>0$, $\mu_{\rm B}$, and $h$ are the g factor, Bohr magneton, and the applied static magnetic field, 
respectively. The first term $A_{1,2}\bm k^2$ of $\epsilon_{1,2}(\bm k)$ denotes the quadratic dispersion, 
while the second term represents the Zeeman energy. The symbol $h_{c1}$ ($h_{c2}$) 
is the critical field of magnons (magnon pairs), and the energy bands are well defined only in the range of $h>h_{c1,c2}$. 
The point is that the Zeeman energy of magnon pairs is twice larger than that of magnons 
because magnon pairs consist of two of down spins. 
From this nature, one can control the distance between two bands by tuning the magnitude of the static field $h$. 
In the main text, we have focused on a tough condition where magnon resonance frequency $\omega_1$ is very close 
to that of magnon pairs, $\omega_2$. The above nature of the Zeeman energy also tells us that 
one can control the difference $\Delta\omega=\omega_2-\omega_1$ by varying the value of $h$.

If the magnetic field $h$ decreases down to $h_{c1}$ ($h_{c2}$), 
$\epsilon_1(\bm k=\bm 0)$ ($\epsilon_2(\bm k=\bm 0)$) becomes negative and 
the magnon (magnon-pair) condensation begins, i.e., the fully polarized state is broken down. 
As we discussed in the main text, the creation and annihilation operators of magnons (magnon pairs) are 
given by $S_{\bm r}^-$ and $S_{\bm r}^+$ ($S_{\bm r}^-S_{\bm r'}^-$ and $S_{\bm r}^+S_{\bm r'}^+$), respectively. 
Therefore, the magnon (magnon-pair) condensation brings the emergence of a transverse magnetization 
$\langle S_{\bm r}^\pm\rangle$ (a spin nematic order $\langle S_{\bm r}^\pm S_{\bm r'}^\pm \rangle$). 
In usual magnets, $h_{c1}$ is larger than $h_{c2}$ or magnon pairs do not exist. 
Thereby, a magnon condensation occurs by lowering the field $h$. 
On the other hand, spin-nematic magnets satisfy $h_{c2}>h_{c1}$ and 
as a result, a magnon-pair condensation and a spin-nematic order can occur just below the critical field $h=h_{c2}$.

\section{Energy Eigenstates for Model (1)}
In this section, we explain the details of energy eigenstates of the frustrated four-spin model (1) we study in the main text.
The Hamiltonian is 
\begin{align}
	{\cal H}_{0}=&\sum_{j=1}^4\left(J_1{\bm S}_j\cdot{\bm S}_{j+1}	+	\varDelta_1{S}_j^z{S}_{j+1}^z\right)	
+\sum_{j=1,2}\left(J_2{\bm S}_j\cdot{\bm S}_{j+2}	+	\varDelta_2{S}_j^z{S}_{j+2}^z	\right)	-	HS_{\rm tot}^z.
\label{eq:4spin}
\end{align}
For the analysis of Eq.~(\ref{eq:4spin}), we introduce three composite spins,
\begin{align}
	\bm{S}_{\rm tot}&=\bm{S}_1+{\bm S}_2+{\bm S}_3+{\bm S}_4,\\
	\bm{S}_\alpha&={\bm S}_1+{\bm S}_3,\\
	\bm{S}_\beta&={\bm S}_2+{\bm S}_4.
\end{align}
Using these spins, we can rewrite the Hamiltonian~(\ref{eq:4spin}) as 
\begin{align}
	{\cal H}_0 =&\frac{J_1}{2}\bm{S}^2_{\rm tot}+\frac{1}{2}(J_2-J_1)(\bm{S}^2_\alpha+\bm{S}^2_\beta)-\frac{3}{2}J_2	
+\frac{\varDelta_1}{2}{S_{\rm tot}^z}^2+\frac{1}{2}(\varDelta_2-\varDelta_1)({S_\alpha^z}^2+{S_\beta^z}^2)-\frac{\varDelta_2}{2}-HS_{\rm tot}^z,
\label{eq:4spin_2}
\end{align}
where we have used the equality ${\bm S}^2=\frac{3}{4}$ for spin-$\frac{1}2$ operator (We use the unit of $\hbar=1$). 
Hereafter, we will ignore the constants $-\frac{3}{2}J_2$ and $-\frac{1}{2}\varDelta_2$ in the Hamiltonian.

First, we consider the SU(2)-symmetric case of $\Delta_{1,2}=0$, in which the Hamiltonian is given by
\begin{equation}
	{\cal H}_0(\Delta_{1,2}=0)\equiv{\cal H}_{\rm su2}=\frac{J_1}{2}\bm{S}^2_{\rm tot}+\frac{1}{2}(J_2-J_1)(\bm{S}^2_\alpha+\bm{S}^2_\beta)
-HS_{\rm tot}^z. 
\label{eq:SU2}
\end{equation}
This model has been analyzed in Ref. [51]. 
The energy eigenvalues and eigenstates are classified by using four quantum numbers,  
$S_{\rm tot}$ (magnitude of total spin $\bm S_{\rm tot}$), $M$ ($z$ component of total spin $\bm S_{\rm tot}$), 
$S_\alpha$ (magnitude of $\bm S_\alpha$) and $S_\beta$ (magnitude of $\bm S_\beta$). Namely, 
the eigenstates are described as $\ket{S_\alpha,S_\beta;S_{\rm tot}, M}$. 
16 eigenstates include a spin-quintet $\ket{1,1;2,M}$, three spin-triplets $\{\ket{1,0;1,M},\,\,\ket{0,1;1,M},\,\,\ket{1,1;1,M}\}$, 
and two spin-singlets $\{\ket{0,0;0,0},\,\,\ket{1,1;0,0}\}$. 
Table~\ref{tab:SU2} represents these wavefunctions with bases $\ket{S_1^z,S_2^z,S_3^z,S_4^z}$, where $S_n^z$ takes 
$\uparrow$ or $\downarrow$, and the corresponding eigen energies $E(S_\alpha,S_\beta;S,M)$. 
We find that two triplets $\ket{1,0;1,M}$ and $\ket{0,1;1,M}$ are degenerate. 
In what follows, for simplicity, 
we re-name $\ket{1,1;2,M}$, $\ket{1,0;1,M}$, $\ket{0,1;1,M}$, $\ket{1,1;1,M}$, $\ket{0,0;0,0}$, and 
$\ket{1,1;0,0}$ as $\ket{2,M}$, $\ket{1,M}_1$, $\ket{1,M}_2$, $\ket{1,M}_3$, $\ket{0,0}_1$, and 
$\ket{0,0}_2$, respectively. 
\begin{table*}[t]
\caption{\label{tab:SU2} Energy eigenstates $\ket{S_\alpha,S_\beta;S,M}$ and eigenvalues $E(S_\alpha,S_\beta;S,M)$ 
of the SU(2)-symmetric model~(\ref{eq:SU2}) with $\Delta_{1,2}=0$. We have eliminated the common constant $-\frac{3}{2}J_2$ from $E(S_\alpha,S_\beta;S,M)$.}
\begin{ruledtabular}
\begin{tabular}{clll}
 Total spin $S_{\rm tot} $ &  $\ket{S_\alpha,S_\beta;S,M}$  &   Representation with bases $|S_1^z,S_2^z,S_3^z,S_4^z \rangle$     & Energy $E(S_\alpha,S_\beta;S,M)$ \\ \hline
 quintet	&	$\ket{1,1;2,2}$		&	$|\uparrow\uparrow\uparrow\uparrow\rangle$	& $J_1+2J_2-2H$ \\
 		&	$\ket{1,1;2,1}$		&	$\frac{1}{2}(|\uparrow\uparrow\uparrow\downarrow\rangle	+	|\uparrow\uparrow\downarrow\uparrow\rangle	+	|\uparrow\downarrow\uparrow\uparrow\rangle	+	|\downarrow\uparrow\uparrow\uparrow\rangle)$
& $J_1+2J_2-H$	\\
 		&	$\ket{1,1;2,0}$		&	$\frac{1}{\sqrt6}(\ket{\uparrow\downarrow\uparrow\downarrow}+\ket{\uparrow\uparrow\downarrow\downarrow}+\ket{\uparrow\downarrow\downarrow\uparrow}+\ket{\downarrow\uparrow\uparrow\downarrow}+\ket{\downarrow\downarrow\uparrow\uparrow}+\ket{\downarrow\uparrow\downarrow\uparrow})$	& $J_1+2J_2$ \\
 		&	$\ket{1,1;2,-1}$	&	$\frac{1}{2}(|\downarrow\downarrow\uparrow\downarrow\rangle	+	|\downarrow\downarrow\downarrow\uparrow\rangle	+	|\uparrow\downarrow\downarrow\downarrow\rangle	+	|\downarrow\uparrow\downarrow\downarrow\rangle)$	& $J_1+2J_2+H$ \\
		&	$\ket{1,1;2,-2}$	&	$|\downarrow\downarrow\downarrow\downarrow\rangle$	& $J_1+2J_2+2H$ \\	
\hline
 triplet	&	$\ket{1,0;1,1}$		&	$\frac{1}{\sqrt2}(|\uparrow\uparrow\uparrow\downarrow\rangle	-	|\uparrow\downarrow\uparrow\uparrow\rangle)$	& $J_2-H$ \\
 		&	$\ket{1,0;1,0}$		&	$\frac{1}{2}(|\uparrow\uparrow\downarrow\downarrow\rangle	-	|\uparrow\downarrow\downarrow\uparrow\rangle	+	|\downarrow\uparrow\uparrow\downarrow\rangle	-	|\downarrow\downarrow\uparrow\uparrow\rangle)$	& $J_2$ \\
 		&	$\ket{1,0;1,-1}$	&	$\frac{1}{\sqrt2}(|\downarrow\uparrow\downarrow\downarrow\rangle	-	|\downarrow\downarrow\downarrow\uparrow\rangle)$	& $J_2+H$ \\	
 		&	$\ket{0,1;1,1}$		&	$\frac{1}{\sqrt2}(|\uparrow\uparrow\downarrow\uparrow\rangle	-	|\downarrow\uparrow\uparrow\uparrow\rangle)$	& $J_2-H$  \\
 		&	$\ket{0,1;1,0}$		&	$\frac{1}{2}(|\uparrow\uparrow\downarrow\downarrow\rangle	+	|\uparrow\downarrow\downarrow\uparrow\rangle	-	|\downarrow\uparrow\uparrow\downarrow\rangle	-	|\downarrow\downarrow\uparrow\uparrow\rangle)$	& $J_2$  \\
 		&	$\ket{0,1;1,-1}$	&	$\frac{1}{\sqrt2}(|\uparrow\downarrow\downarrow\downarrow\rangle	-	|\downarrow\downarrow\uparrow\downarrow\rangle)$	& $J_2+H$ \\	
 		&	$\ket{1,1;1,1}$		&	$\frac{1}{2}(|\uparrow\uparrow\uparrow\downarrow\rangle	+	|\uparrow\downarrow\uparrow\uparrow\rangle	-	|\uparrow\uparrow\downarrow\uparrow\rangle	-	|\downarrow\uparrow\uparrow\uparrow\rangle)$ 
& $-J_1+2J_2-H$	\\
 		&	$\ket{1,1;1,0}$		&	$\frac{1}{\sqrt2}(|\uparrow\downarrow\uparrow\downarrow\rangle	-	|\downarrow\uparrow\downarrow\uparrow\rangle)$	& $-J_1+2J_2$	 \\
 		&	$\ket{1,1;1,-1}$	&	$\frac{1}{2}(|\downarrow\uparrow\downarrow\downarrow\rangle	+	|\downarrow\downarrow\downarrow\uparrow\rangle	-	|\uparrow\downarrow\downarrow\downarrow\rangle	-	|\downarrow\downarrow\uparrow\downarrow\rangle)$	& $-J_1+2J_2+H$	\\	
\hline
 singlet	&	$\ket{0,0;0,0}$		&	$\frac{1}{2}(|\uparrow\uparrow\downarrow\downarrow\rangle	-	|\uparrow\downarrow\downarrow\uparrow\rangle	-	|\downarrow\uparrow\uparrow\downarrow\rangle	+	|\downarrow\downarrow\uparrow\uparrow\rangle)$	& $0$ \\	
		&	$\ket{1,1;0,0}$		&	$\frac{1}{\sqrt3}\left(\ket{\uparrow\downarrow\uparrow\downarrow}-\ket{\downarrow\uparrow\downarrow\uparrow}-\frac{1}{2}(\ket{\uparrow\uparrow\downarrow\downarrow}+\ket{\uparrow\downarrow\downarrow\uparrow}+\ket{\downarrow\uparrow\uparrow\downarrow}+\ket{\downarrow\downarrow\uparrow\uparrow})\right)$	& $-2J_1+2J_2$ \\
\end{tabular}
\end{ruledtabular}
\end{table*}

Next, we consider the anisotropic case with a finite value of $\Delta_{1,2}$ that is the main target of the present study.  
In particular, for simplicity, we focus on the case of $\Delta_1=\Delta_2$, whose Hamiltonian is given by 
\begin{align}
	{\cal H}_0(\Delta_1=\Delta_2)\equiv{\cal H}_1 
=&\frac{J_1}{2}\bm{S}^2_{\rm tot}+\frac{1}{2}(J_2-J_1)(\bm{S}^2_\alpha+\bm{S}^2_\beta)
+\frac{\varDelta_1}{2}{S_{\rm tot}^z}^2-HS_{\rm tot}^z,
\label{eq:4spin_3}
\end{align}
Here, we have already eliminated the constant terms. 
In this case, all the eigenstates of the SU(2) model ${\cal H}_{\rm su2}$ are still those of ${\cal H}_1$. 
We summarize the engenstates and eigenvalues of the anisotropic model ${\cal H}_1$ in Table~\ref{tab:anisotropy}.  
One can see that two triplets $\ket{1,0;1,M}$ and $\ket{0,1;1,M}$ are still degenerate. 

\begin{figure}[t]
\centering
\begin{picture}(250,250)
\hspace{3mm}
\includegraphics[width=7.0cm]{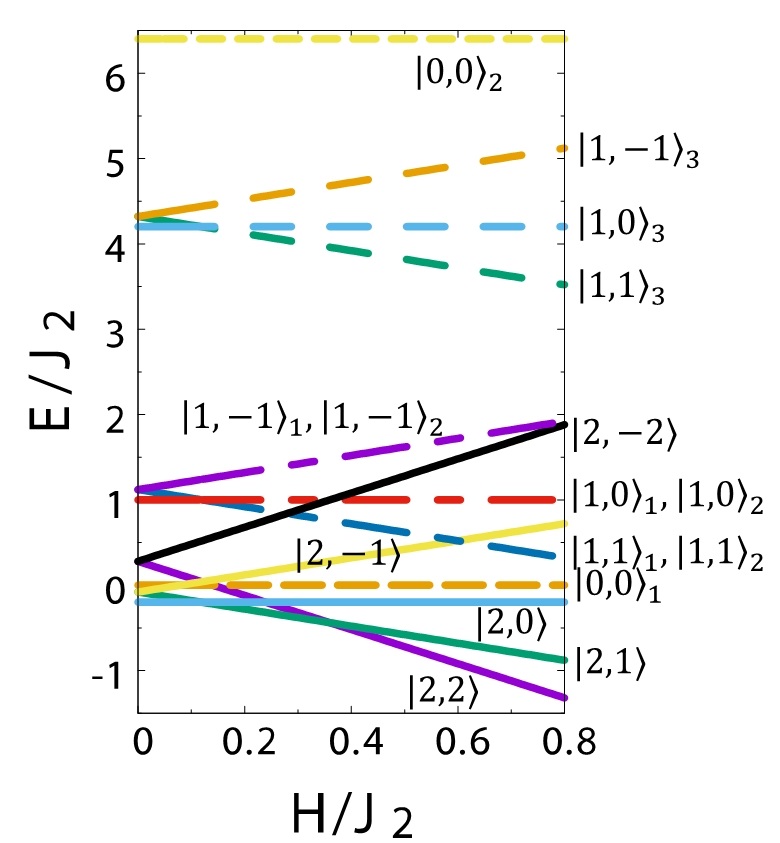}
\end{picture}
\caption{Magnetic-field dependence of eigen energies in the model~(\ref{eq:4spin_3}) with 
$J_1=-2.2$, $J_2=1.0$, and $\Delta_1=\Delta_2=2.4$. 
We have eliminated the common constant $-\frac{3}{2}J_2-\frac{1}{2}\Delta_1$.}
\label{fig:Hdep}
\end{figure}

As we mentioned in the main text, we consider the field-driven fully polarized state. 
Figure~\ref{fig:Hdep} shows the field dependence of energies in the model~(\ref{eq:4spin_3}). 
If we apply a sufficiently strong field $H$, the fully polarized state $|2,2\rangle$ becomes the ground state. 
In the main text, we set the parameters as $J_1=-2.2$, $J_2=1$, $\Delta_1=\Delta_2=0.24$, and $H=0.56$, 
in which the ground state $|\psi_1\rangle$ is fully polarized. Energy eigenstates $\{|\psi_n\rangle\}$ and 
eigenvalues $\{E_n\}$ ($n=1-16$) at this parameter set are given in Table~\ref{tab:value}. 
From this table, we see that the magnon and magnon-pair resonance frequencies, $\omega_1$ and $\omega_2$, 
are estimated as 
\begin{align}
\label{eq:magnon}
	\omega_1&=E_2-E_1=H-\frac{3}{2}\Delta_1=0.2J_2, \\
\label{eq:magnonpair}
	\omega_2&=(E_3-E_1)/2=H-\Delta_1=0.32J_2.
\end{align}

\begin{table}[t]
\caption{\label{tab:anisotropy} Energy engenstates and eigenvalues of the anisotropic model ${\cal H}_1$ 
with $\Delta_1=\Delta_2$. We have eliminated the common constant $-\frac{3}{2}J_2-\frac{1}{2}\Delta_1$ 
from the energies.}
\begin{ruledtabular}
\begin{tabular}{cll}
Total spin	&	Eigen-states $\ket{S,M}_q$	&	Eigen-energies 							\\	\hline
 quintet	&	$\ket{2,2}$		&	$J_1+2J_2+2\Delta_1-2H$			\\
 		&	$\ket{2,1}$		&	$J_1+2J_2+\frac{1}{2}\Delta_1-H$	\\
 		&	$\ket{2,0}$		&	$J_1+2J_2$						\\
 		&	$\ket{2,-1}$		&	$J_1+2J_2+\frac{1}{2}\Delta_1+H$	\\
		&	$\ket{2,2}$		&	$J_1+2J_2+2\Delta_1+2H$		\\	\hline
 triplet	&	$\ket{1,1}_1$		&	$J_2+\frac{1}{2}\Delta_1-H$			\\
		&	$\ket{1,0}_1$		&	$J_2$							\\
		&	$\ket{1,-1}_1$		&	$J_2+\frac{1}{2}\Delta_1+H$			\\	
		&	$\ket{1,1}_2$		&	$J_2+\frac{1}{2}\Delta_1-H$			\\
		&	$\ket{1,0}_2$		&	$J_2$							\\
		&	$\ket{1,-1}_2$		&	$J_2+\frac{1}{2}\Delta_1+H$			\\	
		&	$\ket{1,1}_3$		&	$-J_1+2J_2+\frac{1}{2}\Delta_1-H$	\\
		&	$\ket{1,0}_3$		&	$-J_1+2J_2$						\\
		&	$\ket{1,-1}_3$		&	$-J_1+2J_2+\frac{1}{2}\Delta_1+H$	\\	\hline
 singlet	&	$\ket{0,0}_1$		&	$0$								\\
		&	$\ket{0,0}_2$		&	$-2J_1+2J_2$						\\
\end{tabular}
\end{ruledtabular}
\end{table}

\begin{table}[t]
\caption{\label{tab:value} Energy eigenstate and eigenvalues of the model~(\ref{eq:4spin_3}) 
at $J_1=-2.2$, $J_2=1.0$, $\Delta_1=\Delta_2=0.24$, and $H=0.56$.
}
\begin{ruledtabular}
\begin{tabular}{ll}
 State $\ket{S,M}$				&	Energy	\\	\hline
 $\ket{\psi_1}=\ket{2,2}$		&	$E_1=-0.84$	\\
 $\ket{\psi_2}=\ket{2,1}$		&	$E_2=-0.64$	\\
 $\ket{\psi_3}=\ket{2,0}$		&	$E_3=-0.2$	\\
 $\ket{\psi_4}=\ket{0,0}_1$		&	$E_4=-0.0$	\\
 $\ket{\psi_5}=\ket{2,-1}$		&	$E_5=0.48$	\\
 $\ket{\psi_6}=\ket{1,1}_1$		&	$E_6=0.56$	\\
 $\ket{\psi_7}=\ket{1,1}_2$		&	$E_7=0.56$	\\
 $\ket{\psi_8}=\ket{1,0}_1$		&	$E_8=1.0$	\\
 $\ket{\psi_9}=\ket{1,0}_2$		&	$E_9=1.0$	\\
 $\ket{\psi_{10}}=\ket{2,-2}$		&	$E_{10}=1.4$	\\
 $\ket{\psi_{11}}=\ket{1,-1}_1$	&	$E_{11}=1.68$	\\
 $\ket{\psi_{12}}=\ket{1,-1}_2$	&	$E_{12}=1.68$	\\
 $\ket{\psi_{13}}=\ket{1,1}_3$		&	$E_{13}=3.78$	\\
 $\ket{\psi_{14}}=\ket{1,0}_3$		&	$E_{14}=4.2$	\\
 $\ket{\psi_{15}}=\ket{1,-1}_3$	&	$E_{15}=4.88$	\\
 $\ket{\psi_{16}}=\ket{0,0}_2$		&	$E_{16}=6.4$	\\
\end{tabular}
\end{ruledtabular}
\end{table}

\section{AC Zeeman Coupling}
In this section, we show the matrix form of ac Zeeman coupling driven by applied laser or electromagnetic wave. 
The ac magnetic fields of circularly and linearly polarized laser are respectively represented as 
\begin{equation}
{\bm H}_{c}(t)=A(\cos(\omega t),\sin(\omega t),0),\,\,\,\,\,\,\,
{\bm H}_{\ell}(t)=A(\cos(\omega t),0,0),
\end{equation}
where $A=g\mu_B h_{\rm ac}$ and $\omega$ is the laser frequency. 
Then, the ac Zeeman couplings for circularly and linearly polarized lasers 
are respectively defined as
\begin{align}
{\cal H}_{\rm cp}(t)&={\bm H}_{c}(t)\cdot {\bm S}_{\rm tot}
=\frac{A}{2}\left(\ee^{-i \omega t}S_{\rm tot}^++\ee^{i \omega t}S_{\rm tot}^-\right), \notag\\
{\cal H}_{\rm lp}(t)&={\bm H}_{\ell}(t)\cdot {\bm S}_{\rm tot}=\frac{A}{2}\cos(\omega t)\,\,(S_{\rm tot}^++S_{\rm tot}^-). 
\end{align}
In both cases of circularly and linearly polarizations, the matrix form of the ac Zeeman couplings are written as 
the following type:
\begin{equation}\label{eq:AC}
	{\cal H}_{\rm ac}=\left(\begin{matrix}
{\cal H}^{\rm q}_{\rm ac} & {\bm0} & {\bm0} & {\bm0} & {\bm0}  \\	
{\bm0} & {\cal H}^{\rm t1}_{\rm ac} & {\bm0} & {\bm0} & {\bm0} \\
{\bm0} & {\bm0} & {\cal H}^{\rm t2}_{\rm ac} & {\bm0} & {\bm0} \\
{\bm0} & {\bm0} & {\bm0} & {\cal H}^{\rm t3}_{\rm ac} & {\bm0} \\
{\bm0} & {\bm0} & {\bm0} & {\bm0} & {\cal H}^{\rm s}_{\rm ac}
\end{matrix}\right).
\end{equation}
Here, the first, second, $\cdots$, and 16th lines (or columns) respectively correspond to 
the quintet $\ket{2,M}$, three triplets $\ket{1,M}_1$, $\ket{1,M}_2$, $\ket{1,M}_3$, and two singlets 
$\ket{0,0}_1$, $\ket{0,0}_2$. 

In the circularly polarized case, the quintet part is given by
\begin{equation}\label{eq:quintet}
	{\cal H}^{\rm q}_{\rm ac}= A\left(\begin{matrix}	0&e^{i\omega t}&0&0&0\\
									e^{-i\omega t}&0&\frac{\sqrt6}{2}e^{i\omega t}&0&0\\
									0&\frac{\sqrt6}{2}e^{-i\omega t}&0&\frac{\sqrt6}{2}e^{i\omega t}&0\\
									0&0&\frac{\sqrt6}{2}e^{-i\omega t}&0&e^{i\omega t}\\
									0&0&0&e^{-i\omega t}&0	\end{matrix}\right),
\end{equation}
and the triplet parts are 
\begin{equation}\label{eq:triplet}
	{\cal H}^{\rm t1}_{\rm ac}={\cal H}^{\rm t2}_{\rm ac}={\cal H}^{\rm t3}_{\rm ac} = 
\frac{A}{\sqrt2}\left(\begin{matrix} 
0&e^{i\omega t}&0\\	
e^{-i\omega t}&0&e^{i\omega t}\\	
0&e^{-i\omega t}&0	
\end{matrix}\right).
\end{equation}
The singlet part of the $2\times 2$ matrix ${\cal H}^{\rm s}_{\rm ac}$ is equal to null matrix $\bm 0$.

Similarly, the matrix form of the ac Zeeman term for linearly polarized laser are given by
\begin{equation}\label{eq:quintet_l}
	{\cal H}^{\rm q}_{\rm ac}= A\left(\begin{matrix}	
                                    0&\cos(\omega t)&0&0&0\\
									\cos(\omega t)&0&\frac{\sqrt6}{2}\cos(\omega t)&0&0\\
									0&\frac{\sqrt6}{2}\cos(\omega t)&0&\frac{\sqrt6}{2}\cos(\omega t)&0\\
									0&0&\frac{\sqrt6}{2}\cos(\omega t)&0&\cos(\omega t)\\
									0&0&0&\cos(\omega t)&0	\end{matrix}\right),
\end{equation}
\begin{equation}\label{eq:triplet_l}
	{\cal H}^{\rm t1}_{\rm ac}= {\cal H}^{\rm t2}_{\rm ac}={\cal H}^{\rm t3}_{\rm ac} = \frac{A}{\sqrt2}
\left(\begin{matrix} 
0&\cos(\omega t)&0\\	
\cos(\omega t)&0&\cos(\omega t)\\	
0&\cos(\omega t)&0	
\end{matrix}\right),
\end{equation}
and ${\cal H}^{\rm s}_{\rm ac}=\bm 0$. Employing these matrices and the master equation, we can numerically compute 
the time evolution of the density matrix of the model (\ref{eq:4spin_3}) with ac Zeeman coupling. 

\section{Three- and Four-Magnon Bound States}
In the main text, we have discussed properties of magnon pairs (two-magnon bound states) 
in fully polarized states of spin-nematic magnets. In addition to magnon pairs, 
three- or four-magnon bound states can emerge there, especially, in low-dimensional spin-nematic systems. 
For instance, $J_1$-$J_2$ frustrated ferromagnetic spin chains are shown to possess 
three- or four-magnon bound states as their low-energy excitations in a certain range of $J_1/J_2$ [45].

Here, we shortly discuss whether or not such multiple-magnon bound states disturb 
our proposed method of observing magnon pairs with laser or strong AC field. 
From the conservation law of angular momentum, a process of simultaneous three-photon (four-photon) 
absorption is necessary to create three-magnon (four-magnon) bound states in the polarized states. 
Therefore, a much larger intensity of applied laser is required for the creation compared with that of magnon pairs.  
It indicates that when a magnon-pair resonance takes place with laser, 
resonant peaks of three- or four-magnon bound states can be negligible.

\begin{figure}[t]
\centering
\begin{picture}(250,170)
\includegraphics[width=8.0cm]{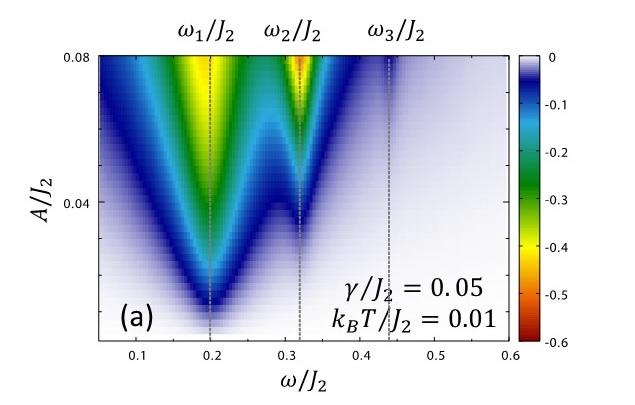}
\end{picture}
\caption{Expanded version of Fig.4(a) of the main text. We set
$J_1=-2.2$, $J_2=1$, $\Delta_1=\Delta_2=0.24$, and $H=0.56$. The resonant points at 
$\omega=\omega_1$, $\omega_2$, and $\omega_3$ respectively correspond to 
single-, two- and three-magnon resonance frequencies.}
\label{fig:3Magnon}
\end{figure}
Even in our nano spin model (1), some resonant peaks of three-magnon (four-magnon) states with 
$S^z_{\rm tot}=-1$ ($S^z_{\rm tot}=-2$) can be observed by tuning the laser frequency $\omega$.  
Figure~\ref{fig:3Magnon} shows a zoom-in version of Fig.4(a). We can observe a small peak at 
$\omega/J_2=\omega_3/J_2=0.44$ at a sufficiently low temperature $k_BT=0.01 J_2$. 
The resonant frequency $\omega_3$ satisfies 
\begin{align}
\label{eq:3magnon}
	\omega_3&=(E_5-E_1)/3=0.44 J_2,
\end{align}
where $E_5$ is the eigen energy of the fifth state $|\psi_5\rangle=|2,-1\rangle$ 
(See Fig.~\ref{fig:Hdep} and Table~\ref{tab:value}). This result clearly shows that the peak at 
$\omega=\omega_3$ corresponds to a three-magnon resonance point. 
The weakness of this peak is consistent with the above expectation. 
Namely, we can neglect the effects of three- four-magnon bound states 
when we observe the magnon-pair resonance.

\end{document}